\documentclass[12pt]{article}
\usepackage{amsfonts,amsmath,amsxtra}

\newcommand{\beq}[1]{\begin{equation}\label{#1}}
\newcommand\eeq {\end{equation}}
\newcommand\bqa {\begin{eqnarray}}
\newcommand\eqa {\end{eqnarray}}
\newcommand\pr {\partial}

\newcommand{\D}{\mathbb{D}}

\newcommand{\A}{\mathbb{A}}
\newcommand{\Z}{\mathbb{Z}}

\begin{document}

\def\t{\theta}
\def\T{\Theta}
\def\w{\omega}
\def\ov{\overline}
\def\a{\alpha}
\def\b{\beta}
\def\g{\gamma}
\def\s{\sigma}
\def\l{\lambda}
\def\wt{\widetilde}

\title{\hfill{ITEP-TH-53/01} \\
\vspace{10mm}
Non-Abelian Structures in BSFT and RR couplings\footnote{Extended
version of the talk presented at 10th Tohwa International Symposium on String
theory which was held in Fukuoka, July 2001.}}


\author{Emil T.Akhmedov\footnote{address:{117259, Moscow, B.Cheremushkinskaya,
25, ITEP, Russia}, email:{akhmedov@itep.ru}}}

\maketitle


\begin{abstract}
In this talk we show that the tachyon annihilation
combined with an approximation, in which string theory
non-commutativity structure is captured by the
algebra of differential operators on space-time,
gives a unified point of view on: non-Abelian structures
on $D$-branes; all lowest energy excitations on
$D$-branes; all RR couplings in type II string theory.
\end{abstract}






{\bf 1.}
The understanding of the relation between large $N$ gauge
theories and string theories is probably one of the most
important problems in the modern theoretical physics \cite{Polyakov}.
Such a relation implies a very interesting
interrelation between the non-Abelian gauge degrees of freedom
of open strings and purely geometrical "gravitational"  degrees of
freedom of closed strings.

  Unfortunately so far we have a poor understanding of the relation in
question mainly due to the lack of the knowledge of
underlying symmetry of string theory. Moreover, it is not clear how the
non-Abelian gauge fields emerge in the case of the near coincident
$D$-branes. As well the clear interpretation  of the closed
strings in open string terms also remains an open question.

Here we would like to address these problems.
To explain how we are going to do that let us remind that open string functionals can be
considered as operators acting on the space of
paths of space-time \cite{WitDoub}. This picture appeared to be quite efficient
in {\bf CS} String Field Theory (SFT). In fact, it helped in
description of tachyon annihilation and $D$-branes as solitons in SFT
\cite{RaSeZw,GrTa}. In this talk we will address similar issues using the
formalism of Boundary SFT (BSFT) \cite{WBI,SBI}. This formalism turned
out to be helpful in verifying the conjectures of \cite{Sen} as it
was demonstrated in \cite{GS1,KuMaMo,KuMaMo1,GS2}. But before dealing with the operators
on the space of paths in BSFT we would like to consider approximation of the string functionals by
differential operators on space-time\footnote{While the consideration of the full non-Abelian structure
of BSFT we postpone for the future.}. We show that the process
of the brane annihilation \cite{Sen} combined with such an approximation leads to the unified description of
backgrounds with various $D$-brane configurations.

The appearance of the differential operators is
also natural from SFT point of view. Configuration
space of the open string theory is roughly given by the maps of
the interval into space-time. If we approximate the strings by
the straight lines (classical limit) the configuration space reduces to the space of
end-points of the intervals in space-time.
Therefore the functionals on configuration space become the
functions of two points and could be interpreted as the kernels of
the integral operators. The expansion around diagonal (image of
the interval is a point) leads naturally to the differential
operators.

  Once we define the approximation in which the non-commutative structure
of the BSFT can be captured by the algebra of differential operators
in space-time, we proceed with the test how the approximation works.
Even in such a trankated version of string theory we derive many "stringy"
phenomena. We observe that all lowest open string excitations can be
obtained as Goldstone type excitations over tachyon profiles corresponding
to the creation of $D$-branes after $D$-anti-$D$-brane ($D-\overline{D}$)
annihilations. In particular, non-Abelian symmetries on $D$-brane bound states
become obvious. Moreover, we give a unified description of all
couplings of the $D-\overline{D}$-brane lowest excitations to the RR massless fields \cite{StCS}.
We choose to study the RR couplings in this context due to their
simple anomalous origin \cite{Moore1}. In fact, they are sensitive to zero
modes and, hence, provide a suitable framework for the discussion of the various open string
backgrounds in our approximation. These couplings were extensively studied
in \cite{BCR}-\cite{SWit}. The picture that emerges could be
considered as an explicit realization of some proposals from
\cite{WitN}.

In an
attempt to provide the unified description of the RR gauge field
couplings for various background brane configurations it is
natural to consider the off-shell interpolation of these
backgrounds. The connection with K-theory uncovered in
\cite{Moore,WitK} leads to the description of the RR-gauge field
couplings in terms of the superconnection \cite{Quillen}. Thus a
natural framework for the universal description of RR couplings
would be in terms of a superconnection in some universal
infinite-dimensional bundle $\mathcal{E}$ over space-time $X$
which provides interpolation between various brane/anti-brane
configurations. This infinite-dimensional bundle $\mathcal{E}$
should be naturally connected with the geometry of $X$ and thus it
is not very surprising to find the "tautological" bundle with
infinite dimensional fiber isomorphic to space of functions on the
base manifolds. This bundle has a rich structure connected with
the action of the differential operators in the fiber. This is in
perfect agreement with the appearance of the differential
operators in the explicit description of the elements of the
K-homolgy groups.

This talk is mainly based on the papers \cite{AkGeSh,GeSh}, but the sections 4 and 7 contain new
considerations and formulas as well.

{\bf 2.}
In this section we explain how
one can capture non-commutative structure of BSFT by
the algebra of differential operators on space-time if the classical
approximation and UV limit are taken with a suitable regularization.

As a model example consider bosonic sigma-model\footnote{All considerations
of the sections 2,3 and 4 are easy to generalize to the case of
superstrings.} on the disk
with  unspecified boundary conditions:

\bqa\label{act1}
S_{2d}=\int_D d\widetilde{X}_\mu d\widetilde{X}^\mu, \quad {\rm where} \quad
\mu = 0,...,25.
\eqa
The action takes extreme values on the configurations:

\bqa
\Delta \widetilde{X}_\mu =0, \quad
*d\,\widetilde{X}_\mu|_{{\partial D}}=0,
\eqa
where $*d$ is the normal derivative $\pr_n$ on the boundary. Thus
on the classical solutions Neumann boundary conditions hold.
On the classical solutions the action $S_{2d}$ is equivalent to:

\bqa\label{lagr1}
S_{2d}^{cl} = \oint_{\pr D} X_\mu \,*d \, X^\mu = \oint_{\pr D} \oint_{\pr D}
d\theta d \theta' X_{\mu}(\theta) H(\theta,\theta')
X^{\mu}(\theta'), \quad {\rm if} \quad \widetilde{X}_\mu |_{\pr D} =
X_\mu(\theta) \label{B}
\eqa
and $\theta$ is some parameterization of the boundary.
Here

$$
H = \sum_k |k| e^{{\rm i} \, k (\theta - \theta')}
$$
is the kernel of the normal derivative operator $*d$ acting on the harmonic functions on the
disk.

It is obvious that the theory (\ref{B}) is equivalent to the following
theory with the additional fields $P_{\mu}$ included:

\bqa\label{lagr}
I_B =\oint_{\pr D} \left( P_{\mu} \, d \, X^{\mu} + P_{\mu} \, * d^{\phantom{\frac12}} P_{\mu}
\right).
\eqa
More precisely, to have the equivalence between the theories (\ref{lagr1}) and (\ref{lagr})
one should not integrate over zero modes of the new fields
$P_{\mu}$ in the functional integral (it may be achieved by
insertion of the delta function $\delta[P_{\mu}(0)]$).
If we consider $X$ and $P$ as dynamical variables in the quantum
mechanics (\ref{lagr}) two-point correlation functions of the fields in this theory are
given by:

\bqa
\left\langle X^{\mu}(\theta)^{\phantom{\frac12}} X^{\nu}(\theta') \right\rangle_{QM}
= \delta_{\mu\nu} G(\theta,\theta')
= \delta^{\mu \nu}\sum_{k\neq 0} \frac{1}{|k|}e^{ik(\theta-\theta')},
\nonumber \\
\left\langle P_{\mu}(\theta)^{\phantom{\frac12}} P_{\nu}(\theta') \right\rangle_{QM} = \delta^{\mu \nu}\sum_{k\neq
 0} \frac{1}{|k|}e^{ik(\theta-\theta')},
\quad
\left\langle X^{\mu}(\theta)^{\phantom{\frac12}} P_{\nu}(\theta')
\right\rangle_{QM}
= \delta^{\mu}_{\nu}\sum_{k\neq 0} \frac{1}{k}e^{ik(\theta-\theta')}.
\eqa
The non-symmetric part  of these correlators  leads to
non-commutativity of the operators $P$ and $X$:

\bqa
\left[X^{\mu}(\theta),\phantom{1^{\frac12}} X^{\nu}(\theta')\right] = 0,
\quad
\left[P_\mu (\theta), \phantom{1^{\frac12}} P_\nu (\theta')\right] = 0,
\quad
\left[P_{\mu}(\theta), \phantom{1^{\frac12}} X^{\nu}(\theta')\right] \propto
\delta_\mu^\nu \, \delta(\theta - \theta'). \label{com}
\eqa
The symmetric part of the correlation functions is given by the
standard Green function:

\bqa
G(\theta,\theta') = \log
|e^{i\theta}-e^{i\theta'}|^2
\eqa
and is singular when the points
coincide. This leads to the necessity to regularize the theory.
 Let us suppose that our regularization prescription
modifies the metric on the boundary as follows:

\bqa\label{reg}
 ||\theta - \theta'|| = |\theta - \theta'| & {\rm when} & |\theta - \theta'| \gg l_R
\nonumber \\
 ||\theta - \theta'|| = 1 & {\rm when} & |\theta - \theta'| \ll l_R,
\eqa
where $l_R$ is a characteristic regularization length. With this
prescription we get for modified Green function $G_R(0)=0$. Note
that this regularization does not modify the "non-commutative"
part of the correlators. Thus we would like to conclude that if we
are interested in the small distance correlation functions only
"non-commutative" parts survive.
In another words, with the
regularization (\ref{reg}) correlation functions at small distances are dominated by
the first term in the lagrangian (\ref{lagr}).
One would like to propose that this is
the way matrix degrees of freedom show up in the sigma model
approach. To show this fact it would be interesting to extract relevant degrees of
freedom explicitly. This is done below.

{\bf 3.} In this section we show that in our approximation string
vertex operators can be represented as differential operators on space-time or more generally
on some auxiliary space. To see this let us
start with Dirichlet boundary condition $\widetilde{X}|_{\partial D} = X(\theta) = Y^1$. It
is easy to construct the boundary operator in conformal field
theory that shifts the Dirichlet boundary conditions $\widetilde{X}|_{\partial
D}= Y^1$ to a new boundary condition $\widetilde{X}|_{\partial D} = Y^1 + \Delta
X(\theta)$:

\bqa\label{use}
V_{\Delta X}=e^{- \int d\theta \, \pr_n X(\theta) \,
\Delta X(\theta)}.
\eqa
In order to realize the process when in the
beginning the string ends
on the $D$-brane $\widetilde{X}|_{\pr D} = Y^1$ then jumps to
$D$-brane $\widetilde{X}|_{\pr D} = Y^2$ and then at
the end jumps back to the $D$-brane $\widetilde{X}|_{\pr D} = Y^1$ we take the
"shift"- function  in the form of step-function:

\bqa
\Delta X = (Y^1 - Y^2) \, \epsilon(\theta|\theta_1,\theta_2), \quad {\rm where}
\eqa
$$\epsilon(\theta|\theta_1,\theta_2)=1 \quad \quad \quad
\theta_1<\theta<\theta_2$$
$$\epsilon(\theta|\theta_1,\theta_2)=0 \quad \quad \quad \quad
{\rm otherwise}.$$
Then using the decomposition
of the scalar field $X=X_+ +X_-$ (on the boundary of the disc)
in terms of positive and negative frequency modes

\bqa
X^{\mu}(\theta) = X_+^{\mu}(\theta)+X_-^{\mu}(\theta)=
\left(\frac{1}{2}X^{\mu}_0+\sum_{k >0}  X^{\mu}_k
 e^{ik\theta}\right) + \left({1 \over 2}X^{\mu}_0+\sum_{k <0}  X^{\mu}_k
 e^{ik\theta}\right)
\eqa
we could represent $V_{\Delta X}$ operator as the product of two "jump"-operators:

\bqa
V_{\Delta X}=\exp\left\{\int^{\theta_1}_{\theta_2} d\theta \, (Y^1-Y^2) \, \pr_n X\right\} =
\exp\left\{\int^{\theta_1}_{\theta_2} d\theta \, (Y^1-Y^2) \, \left[\pr_{t}X_+ -
\pr_{t}X_-\right] \right\} = \nonumber \\ =
\exp\left\{\left[X_+(\theta_1)-X_-(\theta_1)\right]^{\phantom{\frac12}} (Y^1-Y^2)\right\}
\, \exp\left\{-\left[X_+(\theta_2)-X_-(\theta_2)\right]^{\phantom{\frac12}} (Y^1-Y^2)\right\},
\eqa
where $\pr_t = \pr_\theta$ is the tangential derivative to the boundary and we
used the classical equations of motion here.
Therefore, one has the following expression for the "jump"-operator from $Y^i$ to $Y^j$:

\bqa \label{vertex}
V^{ij}\left[X(\theta)\right]
= e^{\frac{(Y^i-Y^j)}{\alpha'} P (\theta)} = e^{\frac{(Y^i-Y^j)}{\alpha'}
\frac{\delta}{\delta X(\theta)}}, \quad {\rm where} \quad P(\theta) = X_+(\theta) -
X_-(\theta).
\eqa
This operator corresponds to the excitation with the mass defined by its
conformal dimension:

\bqa
m^2=\mid Y^i-Y^j \mid ^2/(\alpha ')^2,
\eqa
which is in accordance with the expectations ($m^2 \propto length^2$) for the mass of
the non-Abelian part of the tachyon. In order to get the operators
describing the general states from the spectrum one should
multiply (\ref{vertex}) by the polynomial of the derivatives of $X$ and
the standard exponential factors $e^{ipx}$ responsible for the
non-zero momentum along the $D$-brane.

What is most important for us is that such operators as (\ref{vertex})
respect the D-brane state given under the string
functional integral by the composition of $\delta$-functions
$\delta(X-Y^i)$  where $i=1,...,k$ in the case if there are $k$ $D$-branes (see below). As well
$V^{ij}$ generate the $gl(k)$ algebra when they act on the system of $\delta$-functions
in question \cite{GeSh}.

{\bf 4.} In this section we show how non-Abelian degrees of freedom
on $D$-branes appear after $D-\overline{D}$-brane annihilation if we approximate string functionals
by differential operators on an auxiliary space.

Consider the annihilation of infinite number of $D25$-branes into $k$ $Dp$-branes.
In our approximation the lowest excitations of the $D25$-branes
are described by the following generating functional:

\bqa\label{D25}
Z_{D25} = \left\langle \int \D y^\mu \, \D p_\mu \, \D c^\mu \, \D\bar{c}_\mu\,
\exp\left\{- \oint_{\pr D} d \theta \, \left[{\rm i}^{\phantom{\frac12}} p_\mu \pr_{t} y^\mu + {\rm i} \, \bar{c}_\mu c^\mu +
\right. \right. \right. \nonumber \\ \left. \left. \left.
+ {\rm i} \, A_\mu\left(X|^{\phantom{\frac12}} y,p\right) \, \pr_t X^\mu +
T\left(X|^{\phantom{\frac12}} y,p\right) \right]\right\}
\right\rangle_{BS},
\eqa
where the averaging $\langle ... \rangle_{BS}$ is taken with the bosonic
string functional integral with the action (\ref{act1}).
In this formula we stressed that string fields are differential operators on an auxiliary
space $Y$, parameterized by the coordinates $y^\mu$.
In fact, when there is infinite number of $D25$-branes the space on which Chan-Paton indexes
act becomes more like a Hilbert space of functions on a manifold $Y$ rather than some finite dimensional
vector space. Hence, string functionals are operators acting on this Hilbert space,
i.e. they are functions on $T^*Y$. In "tautological"
situation $Y$ coincides with space-time $X$. It is this situation
we considered above and going to consider below.

 The integral over the fermions $\bar{c}$ and $c$ gives the determinant of
the symplectic form on $T^*Y$ \cite{DustHek}. When $X=Y$ such fermions have a natural
origin in superstring theory as boundary values of NSR fermions (superpartners of $X$)
and their conjugates (superpartners of $P$).
While in bosonic string we have to insert them by hand
according to \cite{DustHek}.

To describe the process of the annihilation of $D25$-brane into
$k$ $Dp$-branes one has to consider the off-shell tachyon
profile as follows \cite{AkGeSh,GeSh}:

\bqa T\left(X|^{\phantom{\frac12}}y,p\right) \propto
\frac{1}{t} |X^\mu - y^\mu|^2 + \frac{1}{s_1} |W(y^a)|^2 + \frac{1}{s_2} |p_\mu|^2,
\label{tachyonprof}
\eqa
where $W(y^a)$ is a polynomial of $y^a$, $a = p+1, ..., 25$
of degree $k$ on zeros of which we are going to localize the
$Dp$-branes; $t$, $s_1$ and $s_2$ are some coupling constants. As is argued in \cite{KuMaMo} the
coupling constant $t$ flows to $0$ in the IR. Hence, the first term
in (\ref{tachyonprof}) describes the localization of space $Y$
on space-time $X$. The second term in (\ref{tachyonprof}) describes
localization on the $Dp$-branes, while the reason of the third one
is explained in the sections 5 and 6, below.

   Let us take the parameters $t,s_1,s_2\to 0$. In case if $A_\mu = 0$
and all zeros of $W$ are separated we obtain \cite{GeSh}:

\bqa\label{delta11}
\lim_{t,s\to 0} \int \D y \, \D p \, \D c \, \D \bar{c} \,
\exp\left\{- \oint_{\pr D} d \theta \, \left[T\left(X|^{\phantom{\frac12}} y,p\right) +
{\rm i} \, p \, \pr_t y + {\rm i} \, \bar{c} \, c \right] \right\}
\propto \sum^k_{i=1} \delta{\left(X_a - Y^i_a\right)}
\eqa
under the string functional integral.
Here $Y^i$ are zeros of $W$. Thus, in this case we obtain just $k$ $Dp$-branes
without any excitations on them.

Now we can deform the background in such a way that the $Dp$-brane system (\ref{delta11}) is
respected in the sense that $Dp$-brane positions are respected. First, we can consider
in (\ref{D25}) $A_\mu(X|y,p) = \left[A^g_m\left(X^m\right)^{\phantom{\frac12}} T^g(y^a,p_a), \,
0, ..., \, 0\right]$, where $m=0,...,p$ enumerate directions along
the $Dp$-branes, $a = p+1, ..., 25$ enumerate directions transversal to the branes
($\mu = (m,a)$) and

\bqa
\left[T^{g_1}, \phantom{1^{\frac12}} T^{g_2}\right] = f^{g_1 g_2 g_3} T^{g_3}, \quad g  = 1,..., k^2 ,
\eqa
i.e. $T^g$ generate the $U(k)$ algebra. The relation of such generators to
the vertex operators $V^{ij}$, $i=1,...,k$ (\ref{vertex}) is described in the section 3 of
\cite{GeSh}. As well
for such a choice of $A_\mu$ the $Dp$-brane system (\ref{delta11}) is
respected, i.e. mapped under the action of the corresponding vertex operators onto
itself. Note that after the limit $s_1,s_2\to \infty$ the group of symplectomorphysms acting
on $T^*Y$ is broken to $U(k)$ \cite{GeSh}.

Second, we can shift $X_a$ via the following operator (we use here (\ref{use})-(\ref{vertex})):

\bqa
\oint_{\pr D} d \theta \, \widetilde{\Phi}^{\tilde{g}}_a(X^m) \, T^{\tilde{g}}(y,p) \,
\frac{\delta}{\delta X^a}
= \oint_{\pr D} d \theta \, \widetilde{\Phi}^{\tilde{g}}_a(X^m) \, T^{\tilde{g}}(y,p) \, \pr_n X^a,
\eqa
for some real functions $\widetilde{\Phi}^g$. In this formula we included
only generators $T^{\tilde{g}}$, $\tilde{g} = 1, ..., k^2 - k$ of $U(k)$,
which do not belong to its Cartan subalgebra. This is done to respect
$Dp$-brane system (\ref{delta11}).

Now after the integration over $y$, $p$, $c$ and $\bar{c}$ via localization
procedure \cite{DustHek}, we obtain:

\bqa\label{Dpbrane}
\lim_{s,t\to 0} Z_{D25} = \left\langle {\rm Tr \, P} \exp\left\{\oint_{\pr
D} d\theta \, \left[{\rm i}\, \hat{A}_m(X^l) \, \pr_t X^m
+ \hat{\Phi}_a(X^l) \, \pr_n X^a\right]\right\}  \right\rangle_{BS},
\eqa
where

\bqa
\hat{A}(X^m) = A^g(X^m) \, t^g, \quad \hat{\Phi}(X^m) =
\widetilde{\Phi}^{\tilde{g}}(X^m) \, t^{\tilde{g}} + {\rm diag}(Y^1, ..., Y^k)
\eqa
and $t^g$, $t^{\tilde{g}}$ are the matrix generators of $U(k)$ in the fundamental
representation corresponding to $T^g$ and $T^{\tilde{g}}$ respectively.
As well  the trace "Tr" in (\ref{Dpbrane}) is taken in the fundamental representation
of $U(k)$. Hence, if all $Y^i$ coincide the full non-Abelian $U(k)$ symmetry is
restored, while if some of $Y^i$ are separated the symmetry is broken
down to a subgroup of $U(k)$.
In conclusion, this formula already describes massless excitations of $k$ $Dp$-branes
rather than of infinite number of $D25$-branes.

{\bf 5.}
 Therefore, in our approximation we can give a geometric unified point of view on
Chan-Paton (non-Abelian) structures in BSFT. Let us describe as well the unified approach
to all open string massless degrees of freedom within our approximation.
We are going to do this on the example of RR couplings to $D$-branes
in type II superstring theory. In this section we just set the notations.

In particular, we describe the RR gauge field couplings with open strings for the
backgrounds with infinite number of $D9-\overline{D9}$-branes and the low
dimensional brane backgrounds arising in the process of the annihilation.
The general coupling of the type IIB $D9-\overline{D9}$-brane fields with the RR fields
is given by

\bqa
\label{coupling} S_{RR}= Q_9\int_{X^{(10)}} \left[C_{RR} \wedge^{\phantom{\frac12}} {\rm
Ch}(\A)\right]_{top}.
\eqa
Here $Q_9$ is the elementary $D9$- and
$\overline{D9}$-brane charge, $X^{(10)}$ is ten-dimensional flat
target space, $C_{RR}$ is defined as the sum of the RR gauge fields of the given
parity (even in IIB theory):

\bqa\label{modes} C_{RR}=\sum_{k}
C_{(2k)}, \quad C_{(2k)} = C_{\mu_1\dots\mu_{2k}}(x) \, dx^{\mu_1}
\wedge \dots \wedge dx^{\mu_{2k}}
\eqa
and the subscript $top$
means that we should integrate the top-dimensional differential
form over $X^{(10)}$; $x^\mu$, $\mu = 0, ..., 9$ are coordinates on $X^{(10)}$.
We use the normalization of RR gauge fields
consistent with the standard definition of the superconnection
used below.

${\rm Ch(\A)}$ is the Chern character of the
superconnection $\A$ \cite{Quillen}, which is constructed from the open string
modes. In fact, the gauge bundles on $D9$- and $\overline{D9}$-branes
may be combined in a $\Z_2$-graded bundle with the superconnection
defined in terms of the gauge field $A_{\mu}$ on the $D9$-brane, the
gauge field $\widetilde{A}_{\mu}$ on the $\overline{D9}$-brane and the
complex tachyon field $T$ corresponding to the lowest energy
excitation of the strings stretched between $D9$- and $\overline{D9}$-branes:

\bqa \label{superconnection} \A = \left(
\begin{array} {c c}
dx^{\mu} \, \nabla^+_{\mu} & T \\ \overline{T} & dx^{\mu} \,
\nabla^-_{\mu}
\end{array}
\right), \nonumber \\ \nabla^+_{\mu} = \partial_{\mu} + {\rm i} \,
A_{\mu}, \quad {\rm and} \quad \nabla^-_{\mu} = \partial_{\mu} +
{\rm i} \, \widetilde{A}_{\mu}.
\eqa
The Chern character of the
superconnection is defined by the standard formula:

\bqa {\rm
Ch}(\A) = {\rm Str} \, e^{- \frac{\A^2}{2\pi}}.
\eqa
Here we use the supertrace "Str" on the matrices acting on $\Z_2$-graded vector
spaces, which is defined as follows. Let $V=V_+\oplus V_-$ be a
$\Z_2$-graded vector space and:

\bqa
\mathcal{O}: V\to V, \quad
\mathcal{O} = \begin{pmatrix}
\mathcal{O}_{++} & \mathcal{O}_{+-} \\ \mathcal{O}_{-+} &
\mathcal{O}_{--}
\end{pmatrix}. \eqa
Here $V_+$ is a bundle on $D9$-brane and $V_-$ is that on
$\overline{D9}$-brane.
Then the  supertrace of $\mathcal{O}$ is defined as:

\bqa
{\rm Str}_{V} \, \mathcal{O} = {\rm Tr}_V \, \tau \, \mathcal{O} = {\rm
Tr}_{V_+} \, \mathcal{O}_{++} - {\rm Tr}_{V_-} \mathcal{O}_{--},
\eqa
where $\tau$ is an operator defining $\Z_2$-structure.
In our case the vector space $V$ is infinite-dimensional
space of functions on an auxiliary space $Y$. Hence, the traces
"Tr" contain integrals over the phase space $T^*Y$:
${\rm Tr}_{\mathcal{H}} ... \propto \int \, dp \, dy$, where
$\mathcal{H} \in V$ is the Hilbert space of functions on $Y$.

Note that to have a well defined trace ${\rm Tr}_{\mathcal{H}}$
the operators should have special property (to
be of the trace class). A particular class of such operators is
given by the expressions $F(p,y)
\propto\delta^{(10)}(p)f(y)$. It's trace is
naturally represented by the integral over $Y$
rather than over $T^*Y$.
This type of operators has a simple qualitative interpretation in
terms of string backgrounds. Let us symbolically denote open
string by the  matrix $K_{y_1,y_2}(x)$ (or more exactly integral
operators $\widehat{K}$ with the kernels $K(x|y_1,y_2)$ acting on the
space of functions on $Y$) where possible end points of the
strings are enumerated by the indexes "$y_1$" and "$y_2$". The trace
invariants naturally correspond to the open strings with
identified ends and thus may be considered as closed strings. Now
let us consider the trace of the operator $\widehat{K}$ with
additional insertion of the operator $\delta(p)$ which projects on the states
with zero eigenvalue of $p$. In case if $X=Y$
it is easy to see that:

\bqa
\delta(p)=\left|p=0^{\phantom{\frac12}}\right\rangle\left\langle \phantom{1^{\frac12}} p=0\right|
= \left(\sum_x |x \rangle \right)\left(\sum_{x'} \langle x'|\right)
\eqa
(note that the identity operator is
$1=\sum_x |x \rangle \, \langle x|$) and we have:

\bqa
{\rm Tr}_{\mathcal{H}} \,
\delta(p) \widehat{K} \propto \sum_{x,x'} K(x,x')
\eqa
We see that in the presence of the operator $\delta (p)$ we should sum over the
positions of the "ends" of the string independently. Thus its
insertion may be interpreted
as the creation of the $D$-brane filling whole space-time.

We have replaced the consideration of the infinite number of
$D$-branes by the consideration of the infinite-dimensional vector
bundles over the space-time $X$ with the fiber --- the space of
functions on the copy $Y$ of $X$. The Lie algebra of
the differential operators acts naturally in the fibers of the
bundle. Note that one may consider infinite dimensional bundles
with the fiber given by the sections of a finite dimensional
(super)-bundle. Below we encounter examples with the space of sections
of spin bundles as a fiber. Having in mind such more general cases treated below we consider
Clifford algebra for the total space $X\otimes T^*Y$:

\bqa
\label{Clif1} \{\gamma_{\mu},\gamma_{\nu}\}=\delta_{\mu \nu} \quad
\{\Gamma_{\mu},\Gamma_{\nu}\}=\delta_{\mu \nu} \quad
\{\widehat{\Gamma}^{\mu},\widehat{\Gamma}^{\nu}\}=\delta^{\mu \nu}
\nonumber \\
\{{\Gamma}_{\mu},\gamma_{\nu}\}=\{\widehat{\Gamma}^{\mu},\Gamma_{\nu}\}=
\{\widehat{\Gamma}^{\mu},\gamma_{\nu}\}=0.
\eqa
The gamma matrices
are defined with respect to the explicit coordinates as follows:
$(x^{\mu},y^{\nu},p_{\rho}) \leftrightarrow
(\gamma_{\mu},\Gamma_{\nu},\widehat{\Gamma}^{\rho})$. Note that
$\Gamma$ and $\widehat{\Gamma}$ (as well as $\gamma$) have a natural
interpretation as zero modes of superpartners of $y$ and $p$ ($X$)
in superstring analog of (\ref{D25}).

{\bf 6.}
Now we give an explicit construction of the superconnection in
the infinite-dimensional bundle in question which leads to the desired description of
RR-couplings. We would like to find such an $\A$ which
after the annihilation of the infinite number of $D9-\overline{D9}$-branes
leads to one $D9$-brane. Similarly to (\ref{tachyonprof}) we have to
obtain something like\footnote{Note that the tachyon vertex on
the $D-\overline{D}$-brane systems
in type II superstring is $\oint |T|^2$ rather than $\oint T$
like in bosonic string \cite{KuMaMo1}.}:

\bqa
\left|T\left(x|^{\phantom{\frac12}} y,p\right)\right|^2 \propto
\frac{1}{t} |x^\mu - y^\mu|^2 + \frac{1}{s} |p_\mu|^2.
\eqa
In fact, the first term is responsible for localization of
$Y$ on $X$ while the second one leads
eventually the trace class operator from ${\rm Ch}(\A)$.
Below we consider localizers like $W$ in (\ref{tachyonprof}) when
we study creation of $Dp$-branes with $p < 9$.

At the same time according to \cite{WitK} we have to consider such a
$T$ which gives a non-trivial map from one chirality spinor
bundles on $X\otimes T^*Y$ to another. Hence, we have to embedd
part of the "Chan-Paton" indexes of $T$ into spinor bundles
on $X\otimes T^*Y$. This means that we take $V = V_+ \oplus V_-$ such
that $V_\pm = s_\pm \otimes S_\pm \otimes \hat{S}_\pm \otimes \mathcal{H}$
and $s_\pm$, $S_\pm$ and $\hat{S}_\pm$ are spinor bundles on $X$, $Y$
and on the fibers of $T^*Y$, respectively. Thus we take the value for the tachyon as follows:

\bqa\label{tach111}
T\left(x|^{\phantom{\frac12}} y, p\right) = \frac{1}{\sqrt{t}} \,
\sigma_{\mu}(x^{\mu}-y^{\mu}) + \frac{1}{\sqrt{s}}\,
\widehat{\Sigma}^{\mu} p_{\mu}.
\eqa
In this formula $\sigma$ and $\widehat{\Sigma}$ are $\sigma$-matrixes
corresponding to $\gamma$-matrixes from (\ref{Clif1}).

As well we choose trivial connections on both $D9$- and $\overline{D9}$-branes
$\nabla = \nabla^+ + \nabla^- = d$. Hence, the corresponding superconnection is

\bqa\label{tremss}
\A_{s,t} = d +
\frac{1}{\sqrt{t}} \, \gamma_{\mu}(x^{\mu}-y^{\mu}) +
\frac{1}{\sqrt{s}} \, \widehat{\Gamma}^{\mu} \, p_\mu
\eqa
This superconnection takes
values in the space of differential operators acting in the
auxiliary space of sections of spinor bundle on $X\otimes T^*Y$. The
$\Z_2$-structure in this case is defined by the chirality. We mean
that with our choice of $V$
the $Z_2$-grading operator is defined by
$\tau = \gamma^{11} \otimes \Gamma^{11} \otimes \widehat{\Gamma}^{11}$,
where $\gamma^{11}$, etc. are ten-dimensional analogs of
$\gamma^5$-matrix in four dimensions.

The square of (\ref{tremss}) is given by the
expression:

\bqa
\A_{s,t}^2=\frac{1}{t}\, \left|x^\mu - y^\mu\right|^2+\frac{1}{s}\,
\left|p_{\mu}\right|^2 +\frac{1}{\sqrt{st}} \,
\gamma_\mu \, \widehat{\Gamma}^\mu + \frac{1}{\sqrt{t}} \,
dx^{\mu} \, \gamma_\mu.
\eqa
The resulting Chern character we are looking for is defined as:

\bqa
{\rm Ch}(\A) = \lim_{t\rightarrow 0} \lim_{s\rightarrow
0}{\rm Ch}(\A_{s,t})
\eqa
Note that here "Str" is taken over the
representation of the full Clifford algebra (\ref{Clif1}) and
includes ${\rm Tr}_{\mathcal{H}}$ as well, i.e. with our
choice of $V$ we have:

$$
{\rm Str} ... = {\rm sp} \, \gamma^{11} \, {\rm Sp} \, \Gamma^{11} \, {\rm
\widehat{Sp}} \, \widehat{\Gamma}^{11} \, {\rm Tr_{\mathcal{H}}} ...,
$$
where by "sp", etc. we define the traces in the spinor part of the bundle $V$
to distinguish them from "Tr" over $\mathcal{H}$.

Thus, using the identities:

\bqa {\rm sp} \left(\gamma^{11^{\phantom{\frac12}}} \gamma_{\mu_1}
\dots \gamma_{\mu_{10}}\right) = (2i)^{5} \epsilon_{\mu_1 \dots
\mu_{10}}
\eqa
and

\bqa
\lim_{t\rightarrow 0} \, \frac{1}{(\pi
t)^{5}} \, \exp\left\{-\frac{|x-y|^2}{t}\right\} =
\delta^{(10)}(x-y), \label{delta}
\eqa
(and similarly for $p$) we obtain:

\bqa
{\rm Ch}(\A) = {\rm
Tr}_{\mathcal{H}} \, \delta^{(10)}\left(p_{\mu}\right)
\, \delta^{(10)} \left(x^\mu-y^\mu\right) = 1.
\eqa
Which means that we have reproduced the
correct coupling with the top-degree RR-form for the $D9$-brane
filling whole space-time:

$$
S_{RR} = Q_9 \, \int_{X^{(10)}} C_{(10)}(x).
$$
This can be considered as the proof that we have made a proper choice of
the background $\A$ (\ref{tremss}).

{\bf 7.} Now let us give a general construction of the RR gauge
field couplings for less trivial $D$-brane backgrounds. We start
with a fixed infinite-dimensional bundle over space-time with
the fiber identified with the space of sections of the spinor
bundle over the base space times the two-dimensional vector space.
We start with the construction of a superconnection corresponding
to parallel $k_+$ $D7$-branes and $k_-$ $\overline{D7}$-branes.
Let $y^a$, $a=8,9$ be coordinates transversal to the $D7$-branes;
$W(y^a)$ is a function defined by the condition that its
critical points are the positions of $D7$- and $\overline{D7}$-branes
in the plane $(y_9,y_8)$: the sign of the Hessian $\pr^2 W$
defines the sign of the RR charge. Consider now the
superconnection:

\bqa\label{D7brane}
\A_{s,t}=d+\frac{1}{\sqrt{t}} \,
\gamma_{\mu}\, \left(x^{\mu} - y^{\mu}\right ) +
\frac{1}{\sqrt{s_1}} \, \Gamma_{a}\, \pr^{a}W(y) +
\frac{1}{\sqrt{s_2}}\, \widehat{\Gamma}^{\mu} \, p_{\mu}
\eqa
In this formula the additional two-dimensional vector space
is interpreted as the representation of the Clifford algebra of
the cotangent space to the transversal space to the $D7$-branes (the notations are in
agreement with (\ref{Clif1})). Note that this representation is
very close to the expression that enter the Supersymmetric Quantum
Mechanic interpretation of Morse theory \cite{Morse}.\footnote{
One could say that we have $N=1$ SUSY Quantum mechanic in the
directions orthogonal to D-brane and $N=\frac{1}{2}$ SUSY Quantum
mechanic in the directions parallel to D-brane.} In fact, the
calculation of the Chern character for this connection in the
limit $t,s_1,s_2 \rightarrow 0$ gives the expression with the
insertion of the projector in the $y$-integral:

\bqa\label{projector}
\det\left[\pr^{a}\, \pr_{b}^{\phantom{\frac12}}W \right] \,
\delta^{(2)}\left[\pr_{a}^{\phantom{\frac12}}W(y)\right]
\eqa
along with $\delta^{(10)}(x-y)$ and $\delta^{(10)}(p)$. Trivial
calculation of ${\rm Tr}_{\mathcal{H}}$ reduces the full expression to the Chern character form
representing $k_+$ $D7$-branes and $k_-$ $\overline{D}7$-branes
($k_{\pm}$ are the numbers of critical points of $W(y)$ with the
positive and negative indexes):

$$
S_{RR} = Q_7 \, \int_{X^{(10)}} \, C_{(8)} (x^\mu) \,\left\{\sum_{i=1}^{k_+} \delta(x^a - Y^{a+}_i) -
\sum_{j=1}^{k_-} \delta(x^a - Y^{a-}_j)\right\},
$$
where $Q_7$ is the elementary $D7$-brane charge and $Y^\pm$ are critical points of $W(y)$
with the positive and negative indexes.

Let us consider now deformations of (\ref{D7brane}) which correspond to
the massless excitations on the $D7$-branes in case if there are no $\overline{D7}$-branes
after the annihilation.
We are interested in deformations which preserve the $D7$-brane system. There are many
gauge equivalent ways to deform the superconnection (\ref{D7brane}),
which are generated by the factor of symplectomorphysms acting
on $T^*Y$ over $U(k)$.
As well there are many deformations which correspond to massive excitations over the
chosen tachyon background and lead to
arbitrary (not necessary parallel) configuration of $D7$-branes.
In this talk we are not going to study the latter kind of deformations.

The simplest deformations which correspond to the massless modes are described by:

\bqa
\A_{s,t}=d + \frac{1}{\sqrt{t}} \,
\gamma_{m}\, \left(x^{m} - y^{m}\right ) + \frac{1}{\sqrt{t}} \,
\gamma_{a}\, \left\{x^{a} - y^{a} - \widetilde{\Phi}^a(y^l|y^a,p_a) \right\} +
\frac{1}{\sqrt{s_1}} \, \Gamma_{a}\, \pr^{a}W(y) + \nonumber \\ +
 \frac{1}{\sqrt{s_2}}\, \widehat{\Gamma}^{m} \, \left\{p_{m} +
{\rm i}^{\phantom{\frac12}} A_m(y^l| y^a,p_a)\right\} +
\frac{1}{\sqrt{s_2}}\, \widehat{\Gamma}^{a} \, p_{a},
\eqa
where $A_m(y^l|y^a,p_a)$, $\widetilde{\Phi}_a(y^l|y^a,p_a)$ are defined as in the section 4
with $m, l = 0, ..., 7$ and $a=8,9$; $y^m$
are coordinates along the directions parallel to the $D7$-brane wold-volume.

The square of this superconnection is

\bqa\label{AA}
\A_{s,t}^2 = \frac{1}{2 s_2} \, \left[\nabla_m, \phantom{1^\frac12}
\nabla_n \right] \, \widehat{\Gamma}^m \, \widehat{\Gamma}^n + \frac{1}{\sqrt{t}} \, dx^{\mu} \,
\gamma_{\mu} + \frac{1}{\sqrt{s_2 t}}\, \left[\nabla_m,
\phantom{1^\frac12} \tilde{\Phi}^a \right] \, \gamma_a \, \widehat{\Gamma}^m +  \nonumber
\\ +  \frac{1}{2t} \, \left[\tilde{\Phi}^{a}, \phantom{1^\frac12} \tilde{\Phi}^{b}\right]
\gamma_{a}\, \gamma_{b} + \frac{1}{t} \, \left|x^{a} - y^a - \tilde{\Phi}^{a}\right|^2
+ \frac{1}{t}\, \left|x^m - y^m\right|^2 + \nonumber \\ + \frac{1}{s_1} \, |\pr_a W(y)|^2 + \frac{1}{\sqrt{s_1
s_2}}\, \pr_a \pr^b W(y) \Gamma^a \widehat{\Gamma}_b + \frac{1}{s_2}\,
|p_a|^2 + \frac{1}{s_2} \left|\nabla_m\right|^2 + ...,
\eqa
where $\nabla_m = p_m + {\rm i}\, A_m(x^l|y^a, p_a)$ and dots stand for
gauge dependent terms which are irrelevant after taking the limit $s_1,s_2\to 0$.

Let us define the symmetric trace "SymTr" following
\cite{Tseytlin,Myers} as the symmetrisation of the usual trace.
Note that:

\bqa\label{simtr}
{\rm Tr} \, e^{A + B + C} = {\rm SymTr} \, e^{A + B + C}
\eqa
Now substituting (\ref{AA}) into (\ref{coupling})
and taking ${\rm Tr}_{\mathcal{H}}$, one finds in the limit $t,s_1,s_2 \to 0$:

\bqa \label{myers2} S_{RR} = Q_7 \,
\int_{M^{(7)}} {\rm SymTr} \,\left[
\exp\left\{ - \frac{1}{2\pi}\left(\hat{F}_{(2)}(x^m) + [\hat{\nabla}_{(1)}(x^m),
\phantom{\frac12} i_{\Phi} ] + [i_{\Phi}, \phantom{\frac12}
i_{\Phi}]\right)\right\} \, C_{RR}(x^m, \, \Phi^a)\right]_{top},
\eqa
where $M^{(7)}$ is the $D7$-brane world-volume,
$\hat{F}_{(2)} = [\hat{\nabla}_m, \, \hat{\nabla}_n] \, dx^m \wedge dx^n$,
$\hat{\nabla}_{(1)} = \hat{\nabla}_m \, dx^m$ and $i_{\Phi}$ is defined by
\cite{Myers}:

$$
i_{\Phi} C_{\dots a} \, dx^a = C_{\dots a} \hat{\Phi}^a.
$$
The last two formulas contain $k\times k$ matrixes $\hat{A}$ and $\hat{\Phi}$
instead of $A(x|y,p)$ and $\tilde{\Phi}(x|y,p)$ as in the section 4.
Note that the expression (\ref{myers2}) coincides with $S_{RR}$ considered in \cite{Myers}.
Thus, we observe that the massless excitations $A$ and $\Phi$ on $D$-branes
appear as Goldstone bosons due to the breaking of the symmetry generated by the algebra
of symplectomorphysms down to $U(k)$ algebra.

{\bf 8.}   It is easy to consider within this framework most general configuration
of parallel $Dp-\overline{Dp}$-brane systems with various $p$ along the
lines of \cite{AkGeSh}. As well it is not hard to consider deformations
of the corresponding superconnections which describe non-parallel
$D$-branes \cite{AkGeSh}.

It would be interesting to apply our procedure to topological string
theories, where the approximation of string non-commutativity structures by
the algebra of differential operators is exact at least for some correlation
functions. As well it would be interesting to generalize all our
considerations to the case of the algebra of differential operators on the space
of paths of space-time.

I am grateful to A.Gerasimov for very valuable discussions and S.Shatashvili for
collaboration. As well I would like to thank organizers of the Tohwa International
Symposium and especially Tada Tsukasa and Hajime Aoki for giving me an
opportunity to present my results and for hospitality.
This work was partially supported by RFBR 01-01-00548 and was commissioned by the AIP.

\end{document}